\documentclass[aps,prc,twocolumn,showpacs,showkeys,nofootinbib,superscriptaddress]{revtex4-2}
\usepackage{bm}
\usepackage{longtable}
\usepackage{graphicx}
\usepackage{amsmath}
\usepackage{amsfonts}
\usepackage{amssymb}
\usepackage{natbib} 
\usepackage{setspace}
\usepackage{xspace}
\usepackage{cancel}
\usepackage{physics}
\usepackage{tensor,braket,color,bbm,slashed}
\usepackage{bbold,dsfont}

\usepackage{newtxtext,newtxmath}
\usepackage{mathrsfs}

\usepackage[normalem]{ulem}

\renewcommand{\sout}{\bgroup \color{red} \ULdepth=-.5ex \ULset}

\begin{document}
\title{  \begin{flushright}
    \rightline{PKNU-NuHaTh-2021-01} 
  \end{flushright}
QCD chiral condensate and pseudoscalar-meson properties in the nuclear medium\\
 at finite temperature}

\author{Parada~T.~P.~Hutauruk}
\email{phutauruk@gmail.com}
\affiliation{Department of Physics, Pukyong National University (PKNU), Busan 48513, Korea}

\author{Seung-il Nam}
\email{sinam@pknu.ac.kr}
\affiliation{Department of Physics, Pukyong National University (PKNU), Busan 48513, Korea}
\affiliation{Center for Extreme Nuclear Matters (CENuM), Korea University, Seoul 02841, Korea}
\affiliation{Asia Pacific Center for Theoretical Physics (APCTP), Pohang 37673, Korea}
\date{\today}

\begin{abstract}
The pion and kaon properties in a nuclear medium at nonvanishing temperature as well as the QCD chiral condensate in the presence of a magnetic field for various baryon densities are studied in the Nambu--Jona-Lasinio (NJL) model with the help of the proper--time regularization (PTR) scheme, simulating a QCD confinement. The density dependent of the quark mass in symmetric nuclear matter is obtained from the quark--meson coupling (QMC) model, which shares the same covariant feature with the NJL model, at quark level. We then analyze the QCD chiral condensates, and dynamical masses for various baryon densities at finite temperature and magnetic field as well as the pion and kaon masses, pion and kaon weak--decay constants, pion--and kaon--quark coupling constants, and wave function renormalization factors for various baryon densities at finite temperature. We find that the QCD chiral condensates suppress with increasing temperature and baryon density and enhance under the presence of a magnetic field, which are consistent with other model predictions. Interestingly, the wave function renormalization factors for the pion and kaon increase with respect to temperature and reduce as the baryon density increases are found.
\end{abstract}

\maketitle
\section{Introduction} \label{intro}

A phase structure of the strongly interacting matter of quantum chromodynamics (QCD) at nonvanishing temperature and density as well as in the presence of the external magnetic field becomes interesting topics in nuclear physics community in the last few years. On the theoretical side, many attempts have been done to study the phase diagram of QCD at nonvanishing temperature and chemical potential~\cite{Fukushima:2010bq,MeyerOrtmanns:1996ea,Fischer18,IFS20,GD20,CQL21,AOHPZ20,SKCL20,CB19,CH19,KS21,IIL19,BGHS19,GF20,TRSW20,FPR19,FCP18,KMOO05,Karsch01,JHABBB20,FDKB06,SK16,NK11,Nam13,BBEFKKSS11,Cao21,SWWY20,EFKS11,LP03,Guenther:2020vqg,SS02}, starting from the low chemical potential and/or high temperature, which is relevant for the early universe phenomena~\cite{Vachaspati91} until the low temperature and/or high chemical potential, which is relevant to the central core of the neutron star or compact star phenomena~\cite{FDGS21,Alford:2007xm}. These regimes have been observed using several various QCD inspired models~\cite{Fischer18,CQL21,SKCL20,CB19,FDKB06,SK16,NK11,Nam13} with different approaches~\cite{SKCL20,CB19,FDKB06,Cao21,SWWY20} as well as lattice QCD~\cite{MeyerOrtmanns:1996ea,Karsch01,BBEFKKSS11,EFKS11,LP03,Guenther:2020vqg} and remarkable progresses have been made until today.

However, besides the progresses have been achieved so far, some problems still remain, namely the sign problem~\cite{KW21,BBDKMSW86,ASSS17}, an uncertain location of the critical end point (CEP)~\cite{FKW08,AMPP12,CFHMP13}, and the chiral imbalance~\cite{FKW08,AS21}, which is related to the chiral magnetic effect (CME), as well as the fixed value of critical transition temperature $T_c$. 
From the lattice simulation side, several studies on lattice QCD simulations~\cite{KW21,BBDKMSW86,ASSS17}, which are the nonperturbative first principle approaches, have been hardly tried to investigate those remaining problems. However, the lattice QCD simulation was only able to compute the phase diagram at higher temperature and very low chemical potential. As mentioned earlier, the lattice QCD still suffers a sign problem at nonvanishing chemical potential ($\mu_B \neq 0$), which relates with the central rapidity of the heavy-ion collisions (HIC). Although, many alternative methods and efforts are developed and tried over last few decades~\cite{Wakayama:2020dzz}, it seems no rigorous method/solution is found yet to solve this sign problem in the lattice  simulation. The phase diagram at finite chemical potential still remains to be outstanding open problem until present. Therefore, many more studies on the in--medium physics phenomena at low and high baryon density as well as temperature and presence magnetic field in the QCD inspired models or other model approaches are extremely required to find the solutions as well as to provide more deeper understanding on properties of QCD in medium and phase structure of strongly interacting matter of QCD.

In the literature, there are so many available studies or calculations on properties of QCD in medium  at finite chemical potential, but not many studies or calculations are performed at finite baryon density~\cite{Cao:2020byn}. In the QMC model, at quark level, it becomes possible to calculate the quark properties as a function of baryon density, which is one of the advantages of the QMC model. In the QMC model, the wave function of the quark inside the nucleon (bag) as well as the nucleon which is solved self-consistently via scalar $\sigma$ and vector $\omega$ fields in nuclear matter. Thus, the coupling constants and other related quantities of the QMC model are determined by reproducing the binding energy $E_B =$ -15.7 MeV at saturation density $\rho_0 =$ 0.16 fm$^{-3}$, which guarantees the stability of model. Such a stability property is not found in the NJL model. However, on one side, the QMC model does not have quark dynamics in the model, meaning we cannot directly calculate the QCD chiral condensate in the QMC model, where it can be straightforwardly calculated in the NJL model. Furthermore, it is also difficult to include the magnetic field and temperature effect in the QMC model, but it is easy to include in the NJL model. With these reasons, therefore, in this work, we used a combine model between the QMC and NJL models, which both models are covariant models and built at quark level, in order to be able to compute the chiral condensate, and other meson properties as a function of baryon density (nuclear medium) at finite temperature and magnetic field.

Experimentally, the ongoing and future experiments for determining the equation of state (EOS) of QCD at higher baryon density such as the Relativistic Heavy Ion Collider (RHIC) at BNL~\cite{KLVW15}, Compressed Baryon Matter (CBM) at FAIR/GSI~\cite{FHKLRRS11} and Nuclotron-based Ion Collider facility (NICA) at JINR~\cite{KKMS20} are planned to run the experiment in order to understand the properties of QCD in medium and the phase structure of strongly interacting matter of QCD. These experimental results will be important to pin down the problems in the QCD strongly interacting matter. Also, it would be useful to discriminate the various theoretical model predictions in the different approaches as well as the lattice QCD simulation results that available in the market. It strongly motivates us to study the properties of quark in nuclear medium at finite temperature and magnetic field in this present work.

In this paper we present a pragmatic approach to analyze the chiral condensate and dynamical mass of the quark as well as the properties of the charged light mesons at finite baryon density $\rho_B$ as well as nonvanishing temperature $T$ using the Nambu--Jona-Lasinio model with the help of the PTR scheme to simulate a QCD confinement~\cite{Hellstern:1997nv}. In addition, we also analyze the quark condensate in the presence of a magnetic field $B$ and at finite $T$. This study may provide an insight understanding on the phase structure of QCD matter at finite $\rho_B$ and $T$, as well as in the presence external $B$ and at finite $T$. The NJL model has been widely and successfully used to study the hadron structure in vacuum~\cite{MYCB05,HCT16,HBCT18} or medium~\cite{HB05,CBT05,HOT18,HMOT19,HT19}, neutron star~\cite{LBT06,WMT15} as well as phase transition~\cite{TBC19}. The density dependence of the quark masses, in the work, are taken from the QMC model in symmetric nuclear matter (Further details on how we determine the in-medium quark mass in the QMC model, see Ref.~\cite{HOT18}) at quark level, as the NJL model. We then connect the QMC and NJL models \textit{via} the density dependence of the quark mass from the QMC model. Next we calculate the chiral condensate, and dynamical mass of the quarks at finite $T$ or nonvanishing external $B$ as well as the meson properties at finite $T$ in the NJL model.



This paper is organized as follows. In Sec.~\ref{sec:njlzero} we briefly present the formulas for the vacuum chiral condensate, dynamical masses for the light and strange quarks, and mass of the charged light meson as well as meson weak--decay constant within the NJL model with the help of the PTR scheme. In Sec.~\ref{sec:njlfinite} we present the in--medium current quark mass as a function of the $\rho_B$ that obtained from the QMC model and the NJL chiral condensate as well as dynamical masses for the light and strange quarks at nonvanishing temperature in the presence of a magnetic field. In Sec.~\ref{results} our numerical results are presented and their implication are discussed. Sec.~\ref{summary} is devoted for a summary and conclusion.
%

\section{Vacuum dynamical mass and chiral condensate}
\label{sec:njlzero}
%
In this section we present the vacuum quark properties in the NJL model at vanishing $T$, finite $\rho_B$, and external $B$ (The details and extensive review to NJL model, see Refs.~\cite{HBCT18,Klevansky92,VW91,HK94}). The NJL model is a chiral effective quark theory of QCD that shares the important features of low energy QCD. Hence, it is very useful tool to use for understanding the nonperturbative phenomena at low energy QCD in particular~\cite{HBCT18}. The interactions of the quark--antiquarks inside mesons are described in the NJL effective Lagrangian with a four--fermion contact interaction, that the gluons interactions are simply absorbed to the coupling constant, reads~\cite{HBCT18}
\begin{eqnarray}
  \label{eq:lagNJL}
  \mathscr{L}_{\rm NJL} & = & \bar{\psi} ( i \slashed{\partial} - \hat{m} ) \psi 
  + G_\pi \left[ (\bar{\psi} \lambda_a \psi )^2
    - (\bar{\psi} \lambda_a \gamma_5 \psi )^2 \right]
  \nonumber \\ &&
  - G_\rho \left[ (\bar{\psi} \lambda_a \gamma^\mu \psi )^2
    + (\bar{\psi} \lambda_a \gamma^\mu \gamma_5 \psi )^2 \right],
\end{eqnarray}
where the quark field is defined as $\psi = (u, d, s)^T$, $\hat{m} = \mbox{diag} (m_u, m_d, m_s)$ denotes the current quark mass matrix, and a sum over $a = 0,\cdot \cdot \cdot, 8$ is implied in Eq~(\ref{eq:lagNJL}), where $\lambda_1 ,\cdot \cdot \cdot, \lambda_{8}$ are the Gell-Mann matrices in flavor space and $\lambda_0 \equiv \sqrt{\frac{2}{3}} \mathds{1}_3$ with $\mathds{1}_3$ being the $3 \times 3$ unit matrix. The four-fermion dimensionful coupling constants are represented by $G_\pi$, and $G_\rho$ in units of GeV$^{-2}$, respectively. In this work, we consider the isospin symmetry case, where $m_u = m_d = m_l$, with $m_l$ is the light quark mass. The solutions to the standard vacuum NJL dressed gap equations are given by
\begin{align}
  \label{eq:propvac1}
  S^{-1}_{l}(p)  &= \slashed{p} - M_{l} +i \epsilon, \nonumber \\
  S^{-1}_{s}(p)  &= \slashed{p} - M_{s} +i \epsilon,
\end{align}
where $l = (u, d)$ and $s$ stands for the light and strange quarks, respectively. The dressed (constituent) quark masses for the light quarks $M_u = M_d$ and strange quark $M_s$ are given by
\begin{eqnarray}
  \label{eq:masNJL}
  M_l &=& m_l - 4 G_\pi \braket{\bar{l}l}, \nonumber \\
  M_s &=& m_s - 4 G_\pi \braket{\bar{s}s},
\end{eqnarray}
where the chiral condensates for the light and strange quarks in the PTR scheme can be defined as
\begin{eqnarray}
  \label{eq:qqNJL}
  \braket{\bar{l}l} &=& -\frac{N_c M_{l}}{4\pi^2} \int_{1/\Lambda_{\rm UV}^{2}}^{1/\Lambda_{\rm IR}^{2}} \frac{d\tau}{\tau^2} e^{-\tau(M_{l}^2)}, \nonumber \\
  \braket{\bar{s}s} &=& -\frac{N_c M_{s}}{4\pi^2} \int_{1/\Lambda_{\rm UV}^{2}}^{1/\Lambda_{\rm IR}^{2}} \frac{d\tau}{\tau^2} e^{-\tau(M_{s}^2)},
\end{eqnarray}
where $N_c = 3$ is the number of the quark color. The dynamical quark masses for the light and strange quarks in the the PTR scheme are expressed by
\begin{align}
  \label{eq:massNJLProp}
  M_l &= m_l^{} + \frac{N_c G_\pi M_l}{\pi^2}
  \int_{1/\Lambda_{\rm UV}^{2}}^{1/\Lambda_{\rm IR}^{2}} \frac{d\tau}{\tau^2} \exp\left(-\tau M_l^2\right) \nonumber \\
  M_s &= m_s + \frac{N_c G_\pi M_s}{\pi^2} \int_{1/\Lambda_{\rm UV}^{2}}^{1/\Lambda_{\rm IR}^{2}} \frac{d\tau}{\tau^2} \exp\left(-\tau M_s^2\right),
\end{align}
where $\Lambda_{\rm UV}^{}$ and $\Lambda_{\rm IR}^{}$ are respectively the ultraviolet (UV) and infrared (IR), which simulates confinement in the NJL model, cutoff parameters. The $\Lambda_{\textrm{IR}}$ ensures the quark production thresholds absence. In this work, the value of the $\Lambda_{\textrm{IR}}$ is set based on the value of QCD limit, $ \Lambda_{\textrm{IR}} = \Lambda_{\textrm{QCD}} = 0.24$ GeV~\cite{HBCT18}. Thus, the $\Lambda_{\rm UV}$ is determined by fitting to the vacuum physical pion mass and weak--decay constant of the pion. Further details on determining the NJL parameters in the NJL model will be fully presented and explained in Sec.~\ref{results}.

The charged light mesons like pion and kaon are realized in the NJL model as dressed quark--antiquark bound states. The pion and kaon properties are determined by the Bethe-Salpeter equation (BSE) in the random phase approximation (RPA). The solution to the BSE in the pseudoscalar meson channel is given by a two-body $t$-matrix that depends on the interaction channel. The reduced $t$-matrices in this channel read
\begin{align}
  \label{eq:tmatrix}
  \tau_\pi^{} (p) &= \frac{-2i\,G_\pi}{1 + 2\,G_\pi\,\Pi_\pi (p^2)}, \\
  \tau_K^{} (p) &= \frac{-2i\,G_\pi}{1 + 2\,G_\pi\,\Pi_K (p^2)}, 
\end{align}
where the bubble diagrams for the corresponding pseudoscalar meson channels are respectively expressed by
\begin{align}
  \label{eq:polK}
& \Pi_{\pi}(k^2) = 2 i N_c \int \frac{d^4p}{(2\pi)^4}\ \mbox{Tr}_D \left[\gamma_5\,S_{l}(p) \gamma_5\,S_{l}
    ( p + k) \right],  \\
& \Pi_{K}(k^2) = 2 i N_c \int \frac{d^4p}{(2\pi)^4}\ \mbox{Tr}_D \left[\gamma_5\,S_{l}(p) \gamma_5\,S_{s}
  (p+k) \right].
\end{align}
From the pole mass position conditions in the corresponding $t$-matrices in Eq.~(\ref{eq:tmatrix}), the meson masses can be determined. The expression for the pion and kaon meson masses are then given by
\begin{align}
  \label{eq:polemass}
  1 + 2\, G_\pi\, \Pi_{\pi} (k^2 = m_{\pi}^2) &= 0,\\
  1 + 2\, G_\pi\, \Pi_{K} (k^2 = m_{K}^2) &= 0.
\end{align}  
The residue at a pole in the $\bar{q}q$ $t$-matrix defines the pion--and kaon--quark coupling constants are given by, respectively
\begin{align}
  \label{eq:couplinconstant}
  g_{\pi q q}^{2} = Z_\pi  &= -\left[ \left.\frac{\partial\, \Pi_{\pi} (k^2)}{\partial k^2}
    \right|_{k^2 = m_{\pi}^2} \right]^{-1}, \\
  g_{K q q}^{2} = Z_K &= -\left[ \left.\frac{\partial\, \Pi_{K} (k^2)}{\partial k^2}
    \right|_{k^2 = m_K^2} \right]^{-1},
\end{align}
where $Z_\pi$ and $Z_K$ are the wave function renormalization constants for the pion and kaon, respectively.

To determine the pion and kaon weak--decay constant, we solve the pion and kaon interacting with the vacuum matrix element $\braket{ 0 \mid \mathcal{J}^a_{\mu,5} (0) \mid \pi (p) }$, where $\mathcal{J}^a_{\mu,5} \equiv \bar{\psi} \gamma_\mu \gamma_5 \lambda^a \psi$ is the axial vector current operator for a flavor quantum number $a$. The pion and kaon weak--decay constants in the PTR scheme are respectively given by
\begin{align}
  \label{eq:decayconNJL}
  f_\pi &= \frac{N_c g_{\pi q q}^{} M_l}{4 \pi^2} \int_0^1 dx\,
  \int_{1/\Lambda_{\rm UV}^{2}}^{1/\Lambda_{\rm IR}^{2}} \frac{d\tau}{\tau} e^{-\tau [ M_l^2- x(1-x) m_\pi^2]},  \\
  f_K &= \frac{N_c g_{Kqq}^{}}{4 \pi^2} \int_0^1 dx\,
  \int_{1/\Lambda_{\rm UV}^{2}}^{1/\Lambda_{\rm IR}^{2}} \frac{d\tau}{\tau} \left[ M_s + x (M_l - M_s)\right] \nonumber \\
  &\times e^{-\tau [ M_s^2 - x(M_s^2 -M_l^2)-x(1-x) m_K^2]}.
\end{align}
Next, the properties of quark in nuclear medium at finite temperature and magnetic field as well as the properties of pion and kaon in nuclear medium at finite temperature will be explained in details in Sec.\ref{sec:njlfinite}.

\section{Finite baryon density and temperature}
\label{sec:njlfinite}
%

In this section we extend the vacuum quark properties as described in Section~\ref{sec:njlzero} to the nonvanishing baryon density and temperature as well as presence of the magnetic field together with the effective constituent quark mass in symmetric nuclear medium obtained from the NJL model augmented with the QMC model. Thus, we present the formulas for the properties of pion and kaon for various baryon densities at finite temperature.

\subsection{Finite baryon density}
\label{qmc}
%

Before describing the temperature and magnetic field dependent of the chiral condensate and dynamical masses of the quarks in the NJL model with the help of the PTR scheme, here we briefly describe how we obtain the effective current quark mass in the QMC model at finite baryon density~\cite{Guichon:2018uew,Saito:2005rv}. In this section, we avoid to repeat the full details of the QMC calculation, therefore, for the interested readers, we refer to Refs~\cite{Guichon:2018uew,Saito:2005rv,HOT18}.

In the QMC model, the effective current quark mass $m_q^*$ at finite baryon density is calculated by $ m_q^* = m_q - V_\sigma^q$, where $m_q$ and $V_\sigma^q$ are the current quark mass, and the scalar potential, respectively. The scalar potential is related with the scalar coupling constant $g_\sigma^q$, and $\sigma$ mean field. Note that in present work, we use $m_q =$ 0.0164 GeV in order to be consistent with the vacuum NJL model, which independently calculated in Ref.~\cite{HOT18}.

In the QMC model, the scalar and vector coupling constants are determined by fitting the binding energy $E/A =$ 15.7 MeV at nuclear saturation density $\rho_0 = 0.15$ fm$^{-3}$ as in Ref.~\cite{HOT18}. Then, we used the obtained coupling constant to calculate the current quark mass in medium and the effective nucleon mass. The compression modulus of the QMC model is taken $K = 281.5$ MeV, and the effective nucleon mass $M_N^{*}$ is estimated to be around 0.80 $M_N$, where $M_N =$ 939 MeV is the free space nucleon mass.

Using the obtained parameters and $m_q^*$, we calculate the in--medium constituent quark mass for the light quark mass of Eq.~(\ref{eq:massNJLProp}) for various baryon densities $\rho_B = [0.00-1.25] \rho_0$ in the NJL model as shown in Table~\ref{tab:model1}.

%
\begin{table}[t]
\caption{The density dependent of the in--medium constituent quark masses calculated in the NJL augmented with the QMC model, which adapted from Ref.~\cite{HOT18}. The units for $M_u^*$ and $M_s^*$ are GeV.
}
\label{tab:model1}
\addtolength{\tabcolsep}{2.8pt}
\begin{tabular}{ccccccc} 
  \hline \hline
  $\rho_B^{} / \rho_0^{}$ & $0.00$ & $0.25$ & $0.50$ & $0.75$ &  $1.00$ & $1.25$  \\
  \hline
  $M_u^{*}$ & $0.400$  & $0.370$ &  $0.339$ & $0.307$ & $0.270$ & $0.207$ \\
  \hline
  $M_s^{*} = M_s$ & $0.611$  & $0.611$ &  $0.611$ & $0.611$ & $0.611$ & $0.611$
  \\ \hline \hline
\end{tabular}
\end{table}

\subsection{Finite Temperature}
%

In this section we present the formulas for the QCD chiral condensate, and constituent quark mass at nonvanishing temperature and presence magnetic field as well as the properties of the pion and kaon at nonvanishing temperature. To take the temperature and presence of the magnetic field into account, we modify the NJL effective Lagrangian in Eq.~(\ref{eq:lagNJL}). It gives
\begin{eqnarray}
  \label{eq:lagT}
  \mathscr{L} &=& \mathscr{L}_{\rm NJL} - \frac{1}{4} F_{\mu \nu} F^{\mu \nu},
\end{eqnarray}
where the $F_{\mu \nu} = \partial_\mu \mathscr{A}_\nu -\partial_\nu \mathscr{A}_\mu$ is the electromagnetic tensor. The covariant derivative is defined as $D_\mu = \partial_\mu - i \hat{\mathcal{Q}} e \mathscr{A}_\mu^{\rm ext}$. The presence of the external magnetic field is considered by introducing homogeneous external magnetic field $B$ coupled to the quarks along the $z$-axis direction, defining $\mathscr{A}_\mu^{\rm ext} = \left( 0, -1/2 B_y, 1/2 B_x , 0 \right)$ in the symmetry gauge~\footnote{In the symmetry gauge, it is written as $A_\mu^{\rm ext} = -\frac{1}{2} \delta_{\mu x} B_y + \frac{1}{2} \delta_{\mu y} B_x$}. Thus, the propagators for the light quarks in the Fock--Schwinger representation~\cite{Schwinger:1951nm,Gusynin:1994xp} is given by 
\begin{eqnarray}
  \label{eqB1}
  \tilde{S}_l (p) &=& -i \int_0^\infty d\mathcal{S} e^{-\mathcal{S}\left[p_{||}^2 + p^2_{\perp} \frac{\tan(|\mathcal{Q}_l B \mathcal{S}|)}{|Q_l B \mathcal{S}|)} + M_l^2\right]} \nonumber \\
  &\times& \Bigg\{ \left[1 - i \gamma^1 \gamma^2 \tanh (|\mathcal{Q}_l B \mathcal{S}|) \right] (M_l -\slashed{p}_{||}) \nonumber\\
  &-& \frac{\slashed{p}_{\perp}}{\cosh^2 (|\mathcal{Q}_l B \mathcal{S}|)} \Bigg\}, \nonumber \\
\end{eqnarray}
and for the strange quark,
\begin{eqnarray}
  \tilde{S}_s (p) &=& -i \int_0^\infty d\mathcal{S} e^{-\mathcal{S}\left[p_{||}^2 + p^2_{\perp} \frac{\tan(|\mathcal{Q}_s B \mathcal{S}|)}{|\mathcal{Q}_s B \mathcal{S}|)} + M_s^2\right]} \nonumber \\
  &\times& \Bigg\{ \left[ 1 - i \gamma^1 \gamma^2 \tanh (|\mathcal{Q}_s B \mathcal{S}|) \right] (M_s -\slashed{p}_{||}) \nonumber \\
  &-& \frac{\slashed{p}_{\perp}}{\cosh^2 (|\mathcal{Q}_s B \mathcal{S}|)} \Bigg\}. \nonumber \\
\end{eqnarray}
Here $\mathcal{S}$ is the proper-time variable, which is the same as $\tau$ in the PTR scheme. The $\mathcal{Q}_u$ and $\mathcal{Q}_s$ are the quark electric charge number of the $l (u,d)$- and $s$-quarks, respectively, where $2/3$ for the $u$-quark and $-1/3$ for the $d$- and $s$-quarks. The $p_{||} = (p_3,p_4)$ and $p_{\perp} = (p_1,p_2)$~\footnote{Here we note that $p_{||}^2 = p_0^2 + p_3^2$ and $p_\perp^2 = p_1^2 + p_2^2$.} are respectively the longitudinal (parallel) and transverse (perpendicular) of the quark momenta to the magnetic field direction. Note that the quark propagator representation in the Fock--Schwinger is consistent with that in the NJL regularization scheme.

The finite temperature effect is incorporated by splitting the fermion four-momentum in the quark propagator in Eq.(~\ref{eq:propvac1}) into $p = (p_4 ,\vec{p}) = (\omega_n , \vec{p})$, where $ p_4 = \omega_n = (2n +1) \pi T$ is the fermionic Matsubara frequency, where $T$ temperature and $n ( = 0, 1, 2, \cdot \cdot \cdot)$ is an integer. In standard convention of the Matsubara formalism, the integral of momentum is simply reformulated by
\begin{eqnarray}
  \int \frac{d^4p}{(2\pi)^4} \rightarrow  \int \frac{dp_4}{2\pi} \frac{d^3\mathbf{p}}{(2\pi)^3}  \rightarrow i T \sum_{n=-\infty}^{\infty} \int \frac{d^3 \mathbf{p}}{(2\pi)^3}.
\end{eqnarray}
With $\mathbf{p}$ is the three--component of the momentum. Replacing the vacuum quark propagator in Eq.~(\ref{eq:propvac1}) with the in--medium quark propagator in Eq.~(\ref{eqB1}) and performing the Matsubara formalism to the chiral condensate, we then obtain the chiral condensate at finite temperature and magnetic field. The compact formula for the chiral condensate of the light quarks at nonvanishing temperature, and presence magnetic field in the PTR scheme is given by
\begin{eqnarray}
  \label{eqqqBT}
  \langle \bar{l} l \rangle &=& - \frac{N_c M_l T}{\pi^{3/2}}\int_{\frac{1}{\Lambda_{UV}^2}}^{\frac{1}{\Lambda_{IR}^2}} \frac{d\tau}{\sqrt{\tau}} e^{-\tau(M_l^2)}\Theta_2 [0,e^{-\tau (2 \pi T)^2}] \nonumber \\
  &\times& |\mathcal{Q}_l B| \coth(|\mathcal{Q}_l B| \tau), \nonumber \\
  &=& - \frac{N_c M_l T}{\pi^{3/2}} |\mathcal{Q}_l B| \int_{\frac{1}{\Lambda_{UV}^2}}^{\frac{1}{\Lambda_{IR}^2}} \frac{d\tau}{\sqrt{\tau}} \frac{e^{-\tau(M_l^2)}\Theta_2 [0,e^{-\tau (2 \pi T)^2}]}{\tanh(|\mathcal{Q}_l B| \tau)}, \nonumber\\
\end{eqnarray}
and for the strange quark,
\begin{eqnarray}
  \langle \bar{s} s \rangle  &=& - \frac{N_c M_s T}{\pi^{3/2}}\int_{\frac{1}{\Lambda_{UV}^2}}^{\frac{1}{\Lambda_{IR}^2}} \frac{d\tau}{\sqrt{\tau}} e^{-\tau(M_s^2)}\Theta_2 [0,e^{-\tau (2 \pi T)^2}] \nonumber \\
  &\times& |\mathcal{Q}_s B| \coth(|\mathcal{Q}_s B| \tau), \nonumber \\
  &=& - \frac{N_c M_s T}{\pi^{3/2}} |\mathcal{Q}_s B| \int_{\frac{1}{\Lambda_{UV}^2}}^{\frac{1}{\Lambda_{IR}^2}} \frac{d\tau}{\sqrt{\tau}} \frac{e^{-\tau(M_s^2)}\Theta_2 [0,e^{-\tau (2 \pi T)^2}]}{\tanh(|\mathcal{Q}_s B| \tau)}, \nonumber \\ 
\end{eqnarray}
where $\Theta_2 [0,e^{-\tau (2 \pi T)^2}]$ is the Jacobi theta function and $1/\tanh(x) = \coth(x)$. Note that the expression for the chiral condensate in Eq.~(\ref{eqqqBT}) in the PTR scheme is compatible with the chiral condensate that separates the vacuum and magnetic field contributions like the magnetic field independent regularization (MFIR) technique~\cite{Ebert:1999ht,Avancini:2019wed}.

Analogous to the chiral condensate, the final expressions for the dynamical masses for the light ($u$ and $d$) quarks at nonvanishing temperature and presence of the magnetic field in the PTR scheme are given by
\begin{eqnarray}
  \label{eqproBT}
  M_u &=& m_u + \frac{N_c G_\pi M_u T}{\pi^{3/2}} |\mathcal{Q}_l B| \nonumber \\
  &\times& \int_{\frac{1}{\Lambda_{UV}^2}}^{\frac{1}{\Lambda_{IR}^2}} \frac{d\tau}{\sqrt{\tau}} \frac{e^{-\tau(M_u^2)}\Theta_2 [0,e^{-\tau (2 \pi T)^2}]}{\tanh(|\mathcal{Q}_l B| \tau)}, \nonumber \\
\end{eqnarray}
for the strange quark,
\begin{eqnarray}
  \label{eqproBT2}
  M_s &=& m_s + \frac{N_c G_\pi M_s T}{\pi^{3/2}} |\mathcal{Q}_s B| \nonumber \\
  &\times& \int_{\frac{1}{\Lambda_{UV}^2}}^{\frac{1}{\Lambda_{IR}^2}} \frac{d\tau}{\sqrt{\tau}} \frac{e^{-\tau(M_s^2)}\Theta_2 [0,e^{-\tau (2 \pi T)^2}]}{\tanh(|\mathcal{Q}_s B| \tau)}.
\end{eqnarray}
Similar to the chiral condensate, the dynamical masses for the light ($l = u,d$) and strange quarks in the PTR scheme as a function of temperature can be also separated into the vacuum ($T \rightarrow 0$) and magnetic contributions ($B = 0$), while for the presence of the external magnetic field, the dynamical mass can be separated by the vacuum ($ B \rightarrow 0$) and thermal contributions ($T = 0$). Again, result for the dynamical masses obtained in Eqs.~(\ref{eqproBT})--~(\ref{eqproBT2}) are consistent with the MFIR result.

Besides calculating the chiral condensate and constituent quark mass at finite temperature and magnetic field, we also compute the polarization insertion for the kaon at finite temperature and absence magnetic field. After performing Feynman parameterization, Wick rotation and Matsubara formalism to the polarization insertion, the final expression for the polarization insertion for the kaon in Eq.~(\ref{eq:polK}) at nonvanishing temperature and absence of a magnetic field in the PTR scheme is obtained,
\begin{eqnarray}
  \label{eq:polKTem}
  \Pi_{K} (p^2) &=& \frac{N_c T}{\pi^{3/2}}  \int_0^1 dx \int_{1/\Lambda_{\rm UV}^{2}}^{1/\Lambda_{\rm IR}^{2}}  \frac{d\tau}{\tau^{3/2}} e^{-\tau[(x^2-x)p^2 + x M_s^2 -M_u^2 (x-1)]} \nonumber \\
  &\times& \left[1 + \tau (2p^2(x-x^2)-xM_s^2 + (x-1) M_u^2 + M_u M_s) \right] \nonumber \\
  &\times& \left[ \Theta_2(0,e^{-\tau(4 \pi^2 T^2)})\right].
\end{eqnarray}

Using the formula in Eq.~(\ref{eq:polKTem}) we can straightforwardly determine the kaon mass at nonvanishing temperature through the vacuum pole mass condition in Eq.~(\ref{eq:polemass}), kaon--quark coupling constant at finite temperature and wave function renormalization at finite temperature in Eq.~(\ref{eq:couplinconstant}).

Using a similar procedure as we calculated the polarization insertion for the kaon, the kaon weak--decay constant at nonvanishing temperature and absence of a magnetic field in the PTR scheme is given by
\begin{eqnarray}
  \label{eq:kaondec}
  f_K &=& \frac{N_c g_{K qq} T}{2 \pi^{3/2}} \int_0^1 dx \int_{1/\Lambda_{\rm UV}^{2}}^{1/\Lambda_{\rm IR}^{2}} \frac{d\tau}{\tau^{1/2}} e^{-\tau[(x^2-x)p^2 + x M_s^2 -M_u^2 (x-1)]}, \nonumber \\
  &\times& \left[ (M_s - M_u)(1-x) + M_u \right] \left[ \Theta_2(0,e^{-\tau(4 \pi^2 T^2)})\right].
\end{eqnarray}

For the pion case, it can be straightforwardly obtained by replacing $M_s \rightarrow M_l$ in the kaon polarization insertion and the kaon decay constant and the $g_{Kqq} \rightarrow g_{\pi qq}$ in the kaon weak-decay constant. Note that here we emphasize again that we only consider the temperature and finite baryon density effects without a magnetic field effect for the meson properties in the present work.

\section{Numerical results}
\label{results}
%

Numerical results for the chiral condensate and constituent mass for the light quarks for various baryon densities at nonvanishing temperature and presence of a magnetic field, and the meson--quark coupling constant, wave function renormalization constant, meson mass, and weak decay constant for various baryon densities at only nonvanishing temperature (without considering a magnetic field) are presented.

For a numerical calculation, we determine the free parameters UV-cutoff $\Lambda_{\rm UV}$, and current quark mass for the up quark $m_u$ by fitting to the physical pion mass $m_\pi=$ 0.140 GeV and pion decay constant $f_\pi=$ 0.093 GeV. We then set the value of the $\Lambda_{\rm IR} \equiv \Lambda_{ \rm QCD} =$ 0.240 GeV and $M_u =$ 0.400 GeV. The $G_\pi =$ 19.043 GeV, $m_u =$ 0.0164 GeV, $M_s =$ 0.611 GeV, and $m_s =$ 0.356 GeV as well as $\Lambda_{\rm UV} =$ 0.645 GeV are obtained in vacuum as in Ref.~\cite{HBCT18}. We then employ these obtained parameters to calculate the properties of the quarks at nonvanishing temperature and magnetic field for various baryon densities, and the meson properties at nonvanishing temperature for various baryon densities.

\begin{figure}[t]
  \centering\includegraphics[width=0.85\columnwidth]{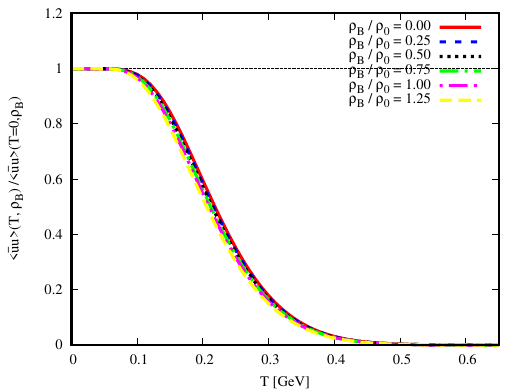}\\
  \centering\includegraphics[width=0.85\columnwidth]{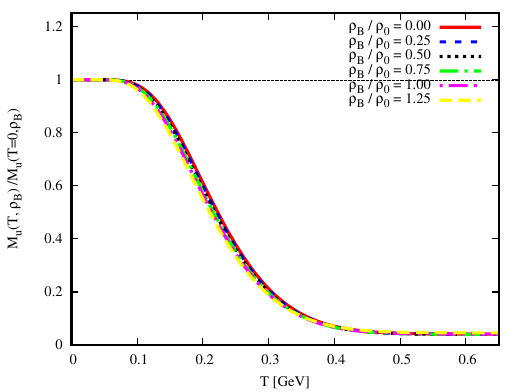}
  \caption{\label{fig1} (Color online) Normalized chiral condensates for the up quark as function of $T$ for various baryon densities (upper panel), and normalized dynamical masses for the up quark as a function of $T$ for various baryon densities (lower panel).}
\end{figure}

\begin{figure}[t]
  \centering\includegraphics[width=0.85\columnwidth]{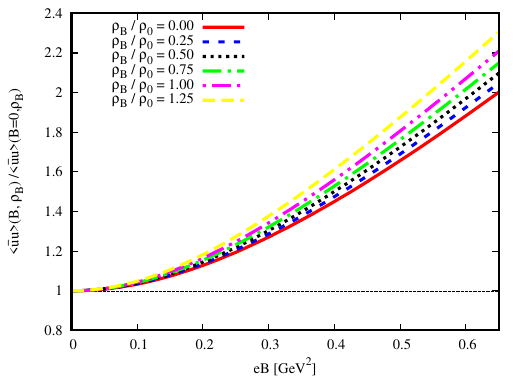}\\
  \centering\includegraphics[width=0.85\columnwidth]{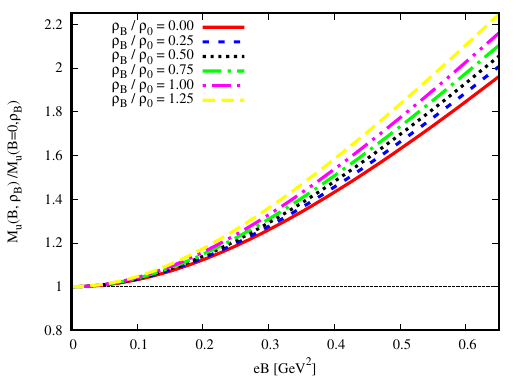}
  \caption{\label{fig2} (Color online) Normalized chiral condensates for the up quark as a function of $eB$ for various baryon densities (upper panel) and normalized dynamical masses for the up quark as a function of $eB$ for various baryon densities (lower panel).}
\end{figure}

\begin{figure}[t]
  \centering\includegraphics[width=0.85\columnwidth]{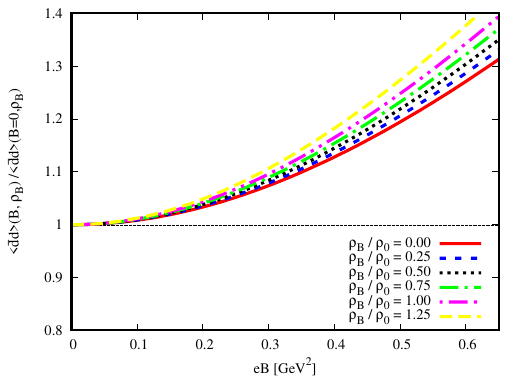}\\
  \centering\includegraphics[width=0.85\columnwidth]{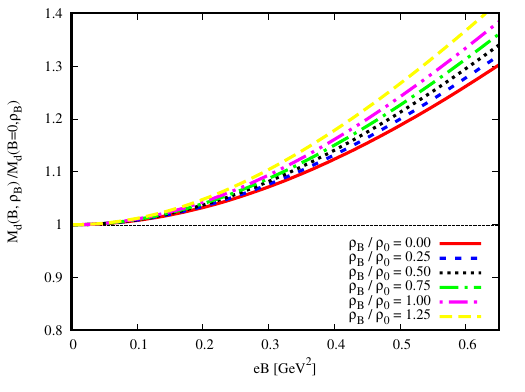}
  \caption{\label{fig3} (Color online) Same as in Fig.~\ref{fig2} but for the down quark.}
\end{figure}

\begin{figure}[t]
  \centering\includegraphics[width=0.85\columnwidth]{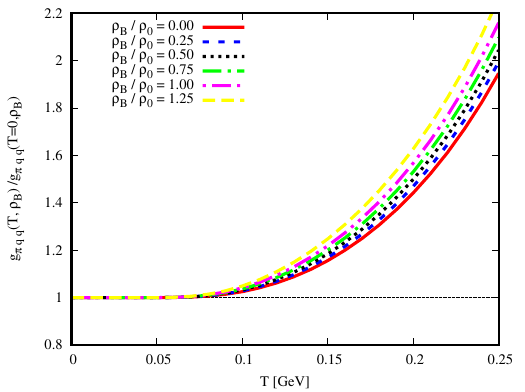}\\
  \centering\includegraphics[width=0.85\columnwidth]{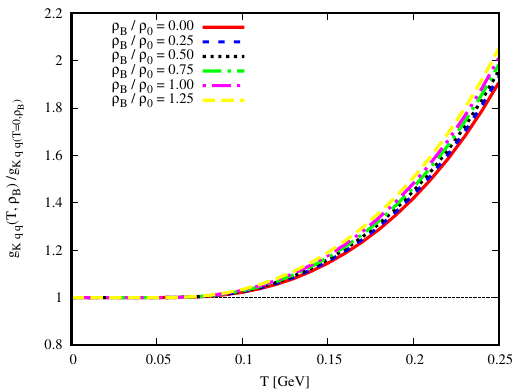}
  \caption{\label{fig4} (Color online) Normalized pion--quark coupling constants as a function of $T$ for various baryon densities (upper panel) and normalized kaon--quark coupling constants as a function of $T$ for various baryon densities (lower panel).}
\end{figure}

\begin{figure}[t] \centering\includegraphics[width=0.85\columnwidth]{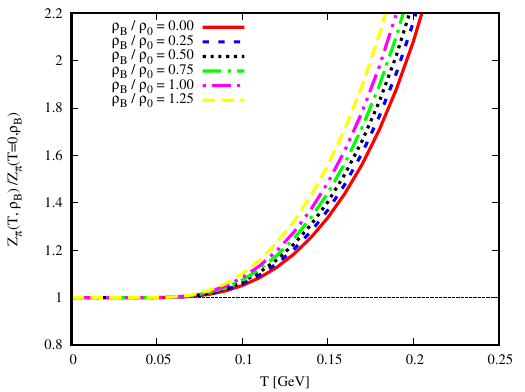}\\\centering\includegraphics[width=0.85\columnwidth]{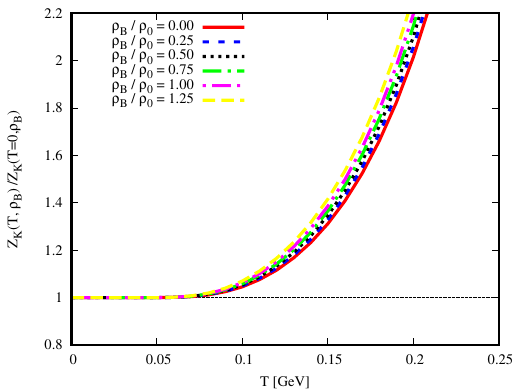}  \caption{\label{fig5} (color online) Same as in Fig.~\ref{fig4} but for the pion (upper panel) and kaon (lower panel) wave function renormalization constants.}\end{figure}

\begin{figure}[t]
  \centering\includegraphics[width=0.85\columnwidth]{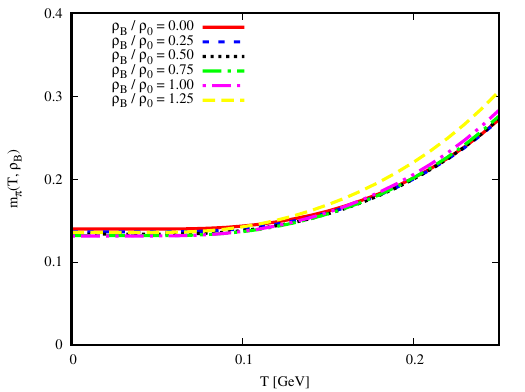}\\
  \centering\includegraphics[width=0.85\columnwidth]{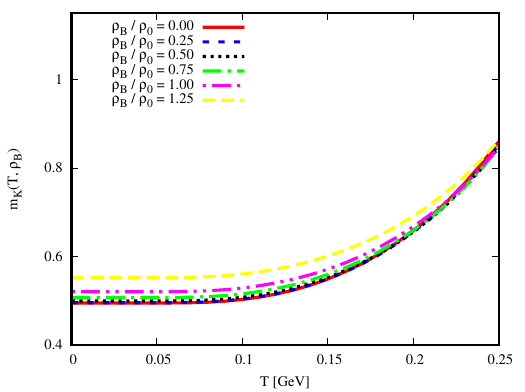}
  \caption{\label{fig6} (Color online) The pion masses for various baryon densities as a function of $T$ (upper panel) and the kaon masses for various baryon densities as a function of $T$ (lower panel).}
\end{figure}

\begin{figure}[t]
  \centering\includegraphics[width=0.85\columnwidth]{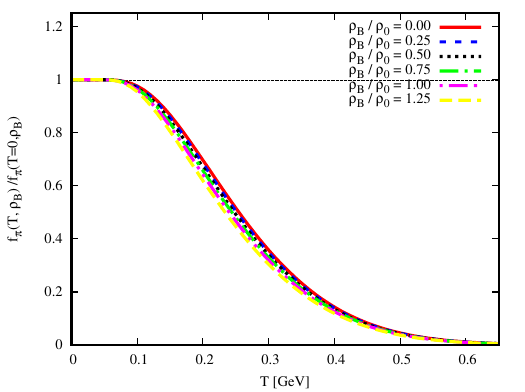}\\
  \centering\includegraphics[width=0.85\columnwidth]{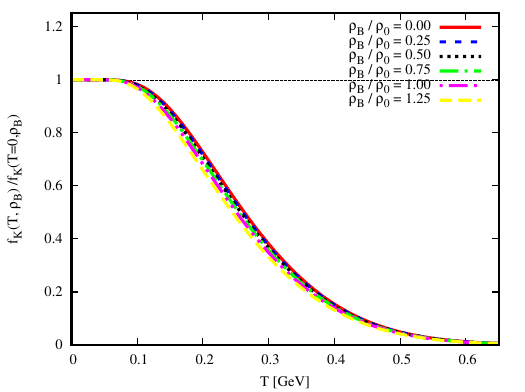}
  \caption{\label{fig7} (Color online) Normalized pion weak--decay constants as a function of $T$ for various baryon densities (upper panel), and normalized kaon weak--decay constants as a function of $T$ for various baryon densities (lower panel).}
\end{figure}

Results for the normalized chiral condensates and dynamical masses for the up quark as a function of $T$ for various baryon densities, which calculated in the NJL model with the PTR scheme, shown in Fig.~\ref{fig1}. Note that, in this work, we consider the isospin symmetry, with $m_u = m_d$. It means that the chiral condensate and dynamical mass for the down quark are taken the same as for the up quark.

The upper panel of Fig.~\ref{fig1} shows the normalized chiral condensates for the up quark as a function of $T$ for various baryon densities. The normalized chiral condensates decrease with increasing the $T$ and baryon densities $\rho_B /\rho_0$ = [0.00, 0.25, 0.50, 0.75, 1.00, 1.25] with $\rho_0$ is the saturation density. Note that the baryon density calculation is adapted from the calculation of Ref.~\cite{HOT18}. Our result for the normalized chiral condensate for the up quark as a function of $T$ for various densities is consistent with the obtained result in Ref.~\cite{Fejos:2017kpq}.

The normalized chiral condensates for the up quark clearly remain constant up to $T =$ 0.1 GeV and they then decrease at $T =$ 0.1-0.3 GeV. For large $T$ the normalized condensates for the up quark for various baryon densities tend to zero. Note that the manifestation of the spontaneously broken of chiral symmetry (SBCS) is represented by the QCD chiral condensate. So, it indicates that the SBCS is restored at around $T = 0.4$ GeV.

The SBCS can also be seen from the transition (critical) temperature $T_c$, which determined \textit{via} the inflection points of the normalized quark condensate curves. We then obtain the $T_c \approx$ 0.19 GeV at $\rho_B = \rho_0$, and $T_c \approx$ 0.20 GeV for $\rho_B/\rho_0 = [0.00, 0.25, 0.50]$. For $\rho_B/\rho_0 =$ 1.25, it gives the $T_c \approx$ 0.18 GeV. The obtained $T_c$ values are consistent with the lattice QCD result in Ref.~\cite{Aoki:2006we}. The $\Delta T_c$ differences are about 10 MeV, where $\Delta T_c$ is the difference between $T_c$ at $\rho_B /\rho_0$ = 1.00 and $T_c$ at $\rho_B/\rho_0 =$ 0.00.

The implication of the SBCS is to generate a dynamical quark mass as shown in the lower panel of Fig.~\ref{fig1}. The SBCS decreases as the baryon density increases. It is quantitatively expected to be lower at higher baryon density.

The normalized dynamical mass for the up quark as a function of temperature for various baryon densities shown in the lower panel of Fig.~\ref{fig1}. Similar with the chiral condensate, the normalized dynamical masses for the up quark for various densities decrease as $T$ increases, as expected. It indicates that the larger contribution of the SBCS to the dynamical mass for the up quark is at around $T =$ 0.0-0.4 GeV.

At higher temperature $T > 0.4$ GeV, the SBCS is not fully restored because of the non--zero current quark mass $m_u \neq 0$, where it has a responsibility for the explicit QCD chiral symmetry breaking. With increasing the baryon densities, the dynamical mass for the up quark decreases, as expected.

The normalized chiral condensates for the up quark as a function of the strength of the magnetic field for various baryon densities depicted in the upper panel of Fig.~\ref{fig2}. The normalized chiral condensates for the up quark increase as the strength of magnetic field increases. This confirms that our result supports the magnetic catalysis (MC) phenomena in a nuclear medium.

This also shows that the SBCS is never restored under influence of the magnetic field. It tends to be larger at higher magnetic field strength. In addition, the normalized chiral condensates for the light up--quark increase as the baryon density increases. This means that the quark condensate of the up quark increases as the value of baryon density increases.

In the lower panel of Fig.~\ref{fig2} we show the normalized dynamical masses for the up quark for various baryon densities as a function of the strength of a magnetic field. The dynamical masses for the up quark increase as the strength of a magnetic field  and baryon density increase, as expected.

We notice that the up and down quarks are separated in the presence magnetic field due to they have different electric charges. We then calculate the normalized chiral condensates and dynamical masses for the down quark for various baryon densities as a function of the  magnetic field as shown in Fig.~\ref{fig3}. Similar features with the up quark, the normalized chiral condensates for the down quark increase with increasing the magnetic field strength, but they are different size in magnitude due to electric charge different. Hence, the enhancement of the chiral condensate for the up quark is higher than that for the down quark. This indicates that the SBCS for the up quark is larger than for the down quark under the presence of a magnetic field.

Also, the chiral condensates for the down quark increase as the baryon densities increase. Consequently, this affects the increasing of the dynamical mass of the down quark as shown in the lower panel of Fig.~\ref{fig3}, as expected. The lower panel of Fig.~\ref{fig3} shows that the normalized dynamical masses for the down quark increase with increasing the baryon density.

The upper panel of Fig.~\ref{fig4} shows the normalized pion--quark coupling constants for various baryon densities as a function of the temperature. The normalized pion--quark coupling constants increase as the temperature increases. This indicates that the interaction strength of the quark inside the pion is more pronounced at higher temperature. The normalized pion--quark coupling constants increase with increasing the baryon density, as expected. In the lower panel of Fig.~\ref{fig4} we show the normalized kaon--quark coupling constants for various baryon densities as a function of temperature. We find that the normalized kaon--quark coupling constants increase as the temperature and baryon density increase. This obviously shows that the kaon--quark coupling constants decrease as the baryon density increases. Comparing with the pion--quark coupling constants, the kaon--quark coupling constants have higher magnitude as in Fig.~\ref{fig4}.

The normalized wave function renormalizations for the pion for various baryon densities as a function of temperature shown in the upper panel of Fig.~\ref{fig5}. We find that the normalized wave function renormalizations for the pion increase as the temperature increases, as expected. The upper panel of Fig.~\ref{fig5} clearly shows that the wave function renormalizations for the pion decrease as the baryon density increases. We do a parameterization for the wave function renormalization of the pion by fitting our NJL calculation data to the polynomial function. We then obtain the expression for the parameterization of the in--medium pion wave function renormalization:
\begin{equation}
  \label{eq:}
  \left( \frac{Z_\pi^{*}}{Z_\pi^{}} \right)  = 1.00 - 0.45 \left(\frac{\rho_B}{\rho_0}\right) + 0.04 \left(\frac{\rho_B}{\rho_0}\right)^2.   
\end{equation}

The lower panel of Fig.~\ref{fig5} shows the normalized kaon wave--function renormalizations for various densities as a function of temperature. Similar to the pion wave--function renormalization case, the kaon wave--function renormalizations increase as the temperature increases. This is followed by the normalized kaon wave--function renormalizations for different values of baryon densities. It also shows that the normalized kaon wave--function renormalizations increase as the baryon density increases. However, it implicitly indicates that the wave function renormalizations for the kaon decrease as the baryon density increases. Using the polynomial function parameterization, the general parameterization formula for the in--medium kaon wave--function renormalization is given by 
\begin{equation}
  \label{eq:}
  \left( \frac{Z_K^{*}}{Z_K^{}} \right) = 1.00 - 0.24 \left(\frac{\rho_B}{\rho_0}\right) + 0.02 \left(\frac{\rho_B}{\rho_0}\right)^2.   
\end{equation}

Using the Goldberger--Treiman relation (GTR), defined as $g_A = g_{\pi NN} f_\pi / M_N$, where $g_A$, $g_{\pi NN}$ and $M_N$ are respectively the iso--vector nucleon weak axial-vector coupling constant, pion--nucleon coupling constant and the nucleon mass. This GTR is originally applied at nucleon level. At quark level, $g_{\pi NN}$ is simply replaced by $g_{\pi qq}$, and $M_N$ is replaced by $M_l$, where $M_l$ is the dynamical mass for the light quark. Straightforwardly, the ratio of the in--medium to vacuum pion--nucleon coupling constant can be expressed by
\begin{eqnarray}
  \label{eq:gtr}
  \left( \frac{g_{\pi NN}^{*}}{g_{\pi NN}^{}} \right) &=& \left(\frac{g_{\pi qq}^{*}}{g_{\pi qq}} \right) \left(\frac{ M_N^{*}}{ M_N^{}} \right) \left(\frac{M_l}{M_l^{*}} \right).
\end{eqnarray}

By plugging the values of the corresponding quantities in Eq.~(\ref{eq:gtr}) from our results, we obtain $ g_{\pi NN}^{*}/g_{\pi NN}^{} \approx $ 0.62 at the saturation density and vanishing temperature. Here, the value of the effective nucleon mass $M_N^*$ is obtained from the QMC model as presented in Section~\ref{qmc}. This findings is rather lower compared to the result obtained in Ref.~\cite{Rakhimov:1998hu} that gives $ g_{\pi NN}^{*}/g_{\pi NN}^{} \approx $ 0.80. However, it is in a good agreement with result in Ref.\cite{Banerjee:1996be} that gives $ (g_{\pi NN}^{*}/M*)/(g_{\pi NN}^{}/M) \approx $ 1.09 at $\rho_B =$ 0.5 $\rho_0$ and our result gives of $ (g_{\pi NN}^{*}/M*)/(g_{\pi NN}^{}/M) \approx $ 0.90.

Result for the pion and kaon masses for different baryon densities as a function of temperature shown in Fig.~\ref{fig6}. The upper panel of Fig.~\ref{fig6} shows the pion mass for different densities as a function of temperature. The pion masses slowly increase as the temperature increases. The pion mass has rather constant value at $T =$ 0.0-0.1 GeV and it starts arising at around $T > 0.1$ GeV. This clearly shows that the pion masses do not dramatically change as the baryon density increases. This result is consistent with the findings in Refs.~\cite{Bernard:1987sx,Kirchbach:1997rk}.

For the kaon masses case, we find that the kaon masses monotonically increase as the temperature increases as shown in the lower panel of Fig.~\ref{fig6}. Also, the kaon masses increase as the baryon density increases, which is consistent with the result of Ref.~\cite{Fejos:2017kpq}.

Figure~\ref{fig7} shows the normalized weak--decay constants for different baryon densities as a function of the temperature. In the upper panel of Fig.~\ref{fig7} we show the normalized weak--decay constants of the pion decrease as the temperature increases. The normalized weak--decay constants for the pion also decrease with increasing the baryon density.

In the lower panel of Fig.~\ref{fig7}, we show the normalized kaon weak--decay constants for different baryon densities as a function of the temperature. Similar to the pion case, we find that the normalized kaon weak--decay constants suppress as the temperature and baryon density increase.

\section{Summary}
\label{summary}
%

As a summary we have studied the QCD chiral condensates for various baryon densities at nonvanishing temperature and presence magnetic field as well as the properties of the light pseudoscalar meson for various baryon densities at nonvanishing temperature in the Nambu--Jona-Lasinio (NJL) model with the help of the PTR scheme, simulating a QCD confinement, which is one of the QCD properties. The density dependent of the current quark mass is obtained from the quark--meson coupling model in symmetric nuclear matter at the quark level, which has the same footing with the NJL model.

For quark properties in nuclear medium at finite temperature and magnetic field, we then calculate the chiral (scalar) condensates, and dynamical quark masses for various baryon densities at finite temperature and presence of a magnetic field. For the meson properties in nuclear medium at finite temperature, we calculate the meson--quark coupling constant, wave function renormalization, kaon mass, and meson weak--decay constant for different baryon densities at nonvanishing temperature.

We find that the normalized quark condensates for the light quark decrease as both temperature and baryon density increase. The decreasing of the normalized QCD chiral condensates for the light quark indicate that our results support the magnetic catalysis (MC) phenomena in nuclear medium. It also shows that the SBCSs are restored at around $T =$ 0.4 GeV. Similar to the chiral condensate case, we find that the normalized dynamical masses for the light quark decrease as both temperature and the baryon density increase.

At presence of a magnetic field, we find the normalized quark condensates for the up quark enhances as the strength of a magnetic field increases, as expected. The normalized up--quark chiral condensates increase with increasing the baryon density. This indicates that the SBCS is not fully restored at presence of a magnetic field. Analogous to the up quark results, the normalized quark condensates for the down quark also increase as the strength of a magnetic field increases and it enhances as the baryon density increases. However, the enhancement of the normalized quark condensates for the up quark are higher compared with that for the down quark. We also find the normalized dynamical masses for the up and down quarks increase as the the strength magnetic field increases and they increase as the baryon density increases.

We find that the normalized pion--quark coupling constants increase as temperature and baryon density increase, as in the upper panel of Fig.~\ref{fig4}. It means that the interaction strength of the pion and quark is more significant at higher temperature. In addition, we find that the normalized kaon--quark coupling constants increase with increasing the temperature. With increasing the baryon density, the normalized kaon--quark coupling constants increase. In connection with the in--medium pion--quark coupling constant through the Goldberger--Treiman relation, we find that the value of $ g_{\pi NN}^{*}/g_{\pi NN}^{} \approx $ 0.62 at the saturation density and $T = 0$.

Result for the normalized wave function renormalizations for the pion and kaon, we find the wave function renormalizations for the pion and kaon increase as temperature increase. The normalized wave function renormalizations for the pion and kaon also increase as the baryon density increases, as expected.

Result for the pion and kaon masses, we find that the pion masses monotonically increase as temperature increases. It almost does not change as the baryon density increases, as expected. The kaon masses increase as both baryon density and temperature increase. We also find the normalized pion and kaon weak--decay constants decrease as both temperature and baryon density increase.

Finally, this study may provide a useful and relevant for a complementary information to understand the properties of the nuclear matter or neutron star, and quark matter as well as QCD phase transition, in particular in the $T-\rho_B/\rho_0$ plane of phase diagrams. To be relevant and more realistic for studying the properties of neutron star or quark matter, another interesting issue related to this study to be included is the rotation effect. Moreover, it would be more interesting and challenging to calculate QCD chiral condensate and meson properties at baryon density (in nuclear medium at quark level) in the consistent model, namely in the NJL model or other more sophisticated models. These issues will be addressed in our future work.

\begin{acknowledgments}
This work was supported by the National Research Foundation of Korea (NRF) grants funded by the Korea government (MSIT) (2018R1A5A1025563 and 2019R1A2C1005697).
\end{acknowledgments}

\end{document}